\documentclass[aps,pre,floats,twocolumn,superscriptaddress,reprint]{revtex4-1}

\usepackage{graphicx,epsfig}

\usepackage{graphics,dcolumn,bm,epic, eepic,float}
\usepackage{amssymb,amsmath,amsfonts,multirow,rotate,color}

\usepackage[table]{xcolor}
\usepackage{flushend}

\begin{document}

\title{Urban Segregation on multilayered transport networks: a random walk approach}

\author{Mateo Neira}
\affiliation{Centre for Advanced Spatial Analysis. University College of London, London, W1T 4TJ, UK}
\affiliation{Alan Turing Institute, British Library, London, NW1 2DB, UK}

\author{Carlos Molinero}
\affiliation{Centre for Advanced Spatial Analysis. University College of London, London, W1T 4TJ, UK}

\author{Stephen Marshall}
\affiliation{Bartlett School of Planning, University College London, London, WC1H 0QB, UK}

\author{Elsa Arcaute}
\affiliation{Centre for Advanced Spatial Analysis. University College of London, London, W1T 4TJ, UK}

\begin{abstract}
We present a novel method for analysing socio-spatial segregation in cities by considering constraints imposed by transportation networks. Using a multilayered network approach, we model the interaction probabilities of socio-economic groups with random walks and L\'evy flights. This method allows for evaluation of new transport infrastructure's impact on segregation while quantifying each network's contribution to interaction opportunities. The proposed random walk segregation index measures the probability of individuals encountering diverse social groups based on their available means of transit via random walks. The index incorporates temporal constraints in urban mobility with a parameter, $\alpha \in [0,1)$, of the probability of the random walk continuing at each time step. By applying this to a toy model and conducting a sensitivity analysis, we explore how the index changes dependent on this temporal constraint. When the parameter equals zero, the measure simplifies to an isolation index. When the parameter approaches one it represents the city's overall socio-economic distribution by mirroring the steady-state of the random walk process. Using Cuenca, Ecuador as a case study, we illustrate the method's applicability in transportation planning as a valuable tool for policymakers, addressing spatial distribution of socio-economic groups and the connectivity of existing transport networks, thus promoting equitable interactions throughout the city. 
\end{abstract}

\maketitle
Cities exist to connect people, influencing human interactions through their structural networks \cite{comunian2011rethinking, barthelemy2016structure, batty2013new}. A successful city maximises face-to-face interactions\cite{comunian2011rethinking}, providing equal opportunities for all inhabitants \cite{pereira2017distributive}. However, understanding the dynamics in cities remains a challenge due to the complexity of urbanisation, social effects, and policy concerns \cite{scott2015nature}. One critical factor influencing these dynamics is urban segregation, which affects socio-spatial interactions within cities.
    
    Urban segregation refers to the spatial separation of different social, ethnic, racial, or economic groups within a city \cite{reardon20022}. It is characterised by the unequal distribution of these groups across neighbourhoods or areas within a city - presenting distinct socio-spatial patterns \cite{massey1988dimensions}. Urban segregation can occur due to various factors, such as historical policies, social preferences, economic disparities, or discrimination \cite{lee2008beyond}. The unequal distribution of population groups in cities can have significant consequences, affecting access to resources, opportunities (both social and economic) while reinforcing social isolation and perpetuating inequalities \cite{louf2016patterns}. Understanding urban segregation is essential for fostering social integration within cities, reducing disparities and promoting more equitable and inclusive places  \cite{louail2017crowdsourcing}. 
    
    Although urban segregation has been extensively studied, there is limited research investigating how coupled transport systems affect urban segregation. To address this gap, the current study integrates concepts from complexity science and complex network analysis, proposing a novel approach to examine urban segregation in relation to transport networks to quantify the likelihood of interactions between different population groups. This study aims to shed light on the intricate connections between urban mobility and segregation, thereby providing insights that can inform more inclusive and equitable urban planning and policies.
    
    Residential segregation can be defined as any pattern in the spatial distribution of population groups that deviates significantly from a random distribution \cite{winship1977revaluation}. This distribution can be a product of social and spatial differentiation, as people have different preferences and resources. People's individual choices can lead to an aggregate outcome that is completely different than what one would expect \cite{Bouchaud2013}. These ideas were formalised by physicists who coined the term \textit{sociodynamics} \cite{weidlich1971statistical} and \textit{sociophysics} \cite{galam1982sociophysics}. Segregation, diffusion, and collective behaviour were explored through this lens with the seminal work of Thomas Schelling (1971) \cite{schelling1971dynamic} and Mark Granovetter (1983) \cite{granovetter1983threshold}. Most statistical methods used to measure segregation found in the literature aim to assess how evenly different population groups are distributed in different areas of the city. 
    
    Although many methods to measure urban segregation had been developed, it wasn't until the work of Massey and Denton (1988) \cite{massey1988dimensions} that they were systematically reviewed and organised. Massey and Denton defined segregation as the degree to which two or more groups live separately from one another. The concept of urban segregation now encompasses various types, including residential, workplace, and experienced segregation. Residential segregation refers to the spatial separation of social groups within neighborhoods, while workplace segregation pertains to the unequal distribution of social groups across occupations and workplaces. Experienced segregation, on the other hand, concerns segregation as experienced through the daily activities people undertake in urban areas. Residential and workplace segregation have long been the focus of urban segregation research, with a growing emphasis on experienced segregation in more recently years with the increase amount of data about people's mobility patterns as captured through location enabled devices \cite{morales2019segregation, moro2021mobility}. 
    
    Segregation has also been a problem explored within network science, specifically in the study in social networks; although the framework is different from residential segregation, it has been extended in interesting ways to spatially embedded networks. In social networks, individuals have a tendency to relate with others who are similar to them across different characteristics; a property known as \textit{homophily} \cite{lazarsfeld1954friendship}, and known within network science as \textit{assortative mixing} \cite{newman2003mixing}. This can have important implications in the information people in the network receive, attitudes they form, and interactions they experience \cite{rodriguez2016overview, shalizi2011homophily}. Measures of assortative mixing can be related to the measures of exposure from residential segregation we discussed previously. Many measures have been proposed by different authors, and in general, they can be divided into two approaches; \textit{descriptive graph statistics} and \textit{spectral graph theory} \cite{rodriguez2016overview}. 
    
    More recent studies have increasingly focused on understanding how segregation manifests itself at different scales \cite{jones2015ethnic, randon2020urban, olteanu2019segregation}. However, the scale at which to study segregation still remains an open question, with most research studying segregation using different methods but mostly working with census track data and their adjacencies to derive segregation measures. Underlying these studies there is an assumption that all persons sharing a tract, whether they are located in the centre of the tract or towards the periphery, have equal proximity to residents outside the tract as well as being equally proximate to everyone within its boundaries. This assumption stems from treating tracts as as spatially homogeneous irrespective of their relative distances and the connectivity patterns provided by the street network and additional transport networks that might be present. In our work we address this problem by providing a method that explicitly takes into account a city's connectivity structure by modelling all modes of transport as a multilayered network.
    
    The proposed method for quantifying segregation using multilayer networks and random walks takes into account the heterogeneity of the connectivity different groups of people have dependent of where they are located in a city. Segregation is understood as unequal opportunities for encounters, measured as the lack of \textit{exposure} between different population groups. To measure the lack of exposure between groups their spatial relationships must be taken into account, as well as the constraints that the available transport networks impose on the opportunities for encounters. The paper is structured as follows: we first provide a brief overview of similar studies that have used random walks to measure segregation. Then we introduce the methodology and apply the method to a toy model in order to explore the sensitivity of the parameters. Finally, we apply the framework to study segregation in the city of Cuenca, Ecuador, and we show how the measure can be used to assess the impact of transport infrastructure on segregation.

\section*{Related works}
    Random walk models, utilised across diverse fields like biology, physics, chemistry, economics, and computer science, are significantly influential due to their simplicity. In these models, a 'walker' or moving item explores an area randomly, without any reference to its past paths. However, despite the seeming simplicity, the subsequent dynamics related to the walker’s random movements can be unexpectedly complex. Thus, this underlines the contradiction where the simplicity of a random walk model hides the potential complexity of its results.

    Recently, the interest in random walks has shifted towards the analysis of complex systems via networks. This study often involves random walk methods which enable transitions from any given node to its nearest network neighbours. Such methods are particularly utilised in multilayered networks and could be beneficial in a myriad of applications, including epidemic propagation, social network dynamics, information analysis, and human mobility, among others.
    
    A random-walk based approach to measure residential segregation was introduced by Ballester and Vorsatz (2014) \cite{ballester2014random}, providing an eigenvector-based centrality measure that captures the probability of two individuals from different social groups meeting. Extending the use of random walk-based methods to measure segregation, Sousa and Nicosia (2022) \cite{sousa2022quantifying} proposed a family of non-parametric measures of spatial segregation that use the statistic of the trajectories of random walks on graphs. The research uses the distribution of class coverage times, the expected number of steps required by a random walk to visit a certain fraction of all social groups present in an urban area. By defining a null-model the proposed measure can account for size difference across different urban systems.
    
    In this work we developed a novel approach that considers the city as a multilayered network, incorporating all available transport systems, and combining socio-economic data to measure segregation. Our method utilises a modified random-walk-based segregation index \cite{ballester2014random}, capturing interaction probabilities of diverse population groups while accounting for spatio-temporal constraints imposed by the city's interconnected transport network. This index addresses limitations of previous purely spatial methods by explicitly assessing the impact of transport infrastructure on potential social interactions across various city areas, thereby bridging the gap between spatial and social segregation within urban environments.
    
    Multilayered networks are networks that have multiple layers or types of connections between the same set of nodes. These types of networks allow for a more comprehensive view of the relationships between nodes by looking at multiple types of relationships between them. They can provide insight into the interconnectedness of different social, economic, or biological systems. multilayered networks can be used to analyse the spread of information, the spread of disease, and other complex phenomena. In our work we model the coupling of transport systems as a multilayered network, where each transport network is modelled as a layer connecting different areas in a city.

\section*{Methods}

    \subsection*{Random walk segregation on multilayered networks}
    
        We can study the structure and interactions in urban systems through their networks. In particular, multilayered networks allow us to capture the coupling of multiple transport systems, and by doing so it is able to better capture the spatio-temporal constraints they impose on different places and can be used to measure urban segregation at a fine spatial resolution. 
        
        Here we set a formal description of the independent transport network graphs and their inter-modal coupling to create the multilayered network. We describe the assignment of socio-economic variables to the multilayered network, as well as the measures that will be used to assess change in urban structure and segregation.
        
        Each transport network is modelled as a digraph in their \textit{primal} representation, where each intersection or station is modelled as a node, and their connections - such as streets, routes, or transport lines - are modelled as temporally weighted links. All the transport networks are represented by an ordered list of networks, $\vec{G}$ given by:
            \begin{equation}
            \vec{G} = (G_{1}, G_{2},...,G_{i},...G_{M})
            \end{equation}
        where $M$ indicates the total number of transport modes available in the city, and $G_{i}=(N_{i}, L_{i}, w_{i})$ is the transport network $i$. $N_{i}$ is a set of nodes, $L_{i}$ is a set of links, and $w_{i}$ is a function that takes a link and returns a weight ($w_{i}: L_{i} \rightarrow \mathbb{R}$) equal to the travel time between the links of the node in minutes. In the case of street networks, this is calculated using an average walking speed of $5km/h$.
        
        Additionally, we define the coupling of the different transport networks through a $M \times M$ list of bipartite networks $\mathcal{G}_{i,j} = (N_{i}, N_{j}, L_{i, j}, w_{i,j})$ for each $i < j$ and $ i,j \in \{1,2...M\} $. $\mathcal{G}$ indicates the bipartite network with node sets $N_{i}$ and $N_{j}$ and the link set $L_{i, j}$. The links of the network $\mathcal{G}_{i, j}$ are called interlinks and connect the nodes of layer $i$ to nodes of a different layer $j$ and $w_{i,j}$ is a function that takes a link and returns a weight ($w_{i,j}: E_{i,j} \rightarrow \mathbb{R}$) equal to the transfer time in minutes between two transport layers.
        
        Finally, the multilayer network $\mathcal{M}$ is given by the triple:
            \begin{equation}
            \mathcal{M} = (Y, \vec{G},\mathcal{G})
            \end{equation}
        where $Y$ indicates the set of layers $Y=\{i|i \in \{1,2,...,M\}\}$ for each transport network. 
        This multilayered network can be described by a supra-adjacency matrix $A_{\mathcal{M}}$\cite{kivela2014multilayer} and the corresponding  time-weighted supra-adjacency matrix $T_{\mathcal{M}}$. 
        
        The minimum number of transport networks a city can have is one, corresponding to the street network. Socio-economic data needs to be incorporated into the nodes of the street network to measure urban segregation using the multilayered model. Formally, this is captured by classifying each individual from a total set $N = \{1,2,...,n\}$ of $n$ individuals in a city, into different socio-economic groups $B$. Let $n_{b,i}$ be the number of individuals of group $b \in B$ that will start a journey through the city in the multilayered nodes $i \in N_{\mu}$, where $\mu$ denotes the layer, and in particular $\mu=1$ corresponds to the street network one.
        The number of individuals who belong to a group $b$ is $n_{b} = \sum_{i} n_{b,i}$, and the  number of individuals who will start their journey from node $i$ is $n_{i} = \sum_{b \in B} n_{b,i}$. The column vectors $c_{b} = (\frac{n_{b,i}}{n_{i}})_{i}$ and $d_{b} = (\frac{n_{b,i}}{n_{b}})_{i}$ are referred to as the vectors of group concentrations and group densities, respectively. 
        
\begin{figure}
            \centering
            \includegraphics[width=1\linewidth]{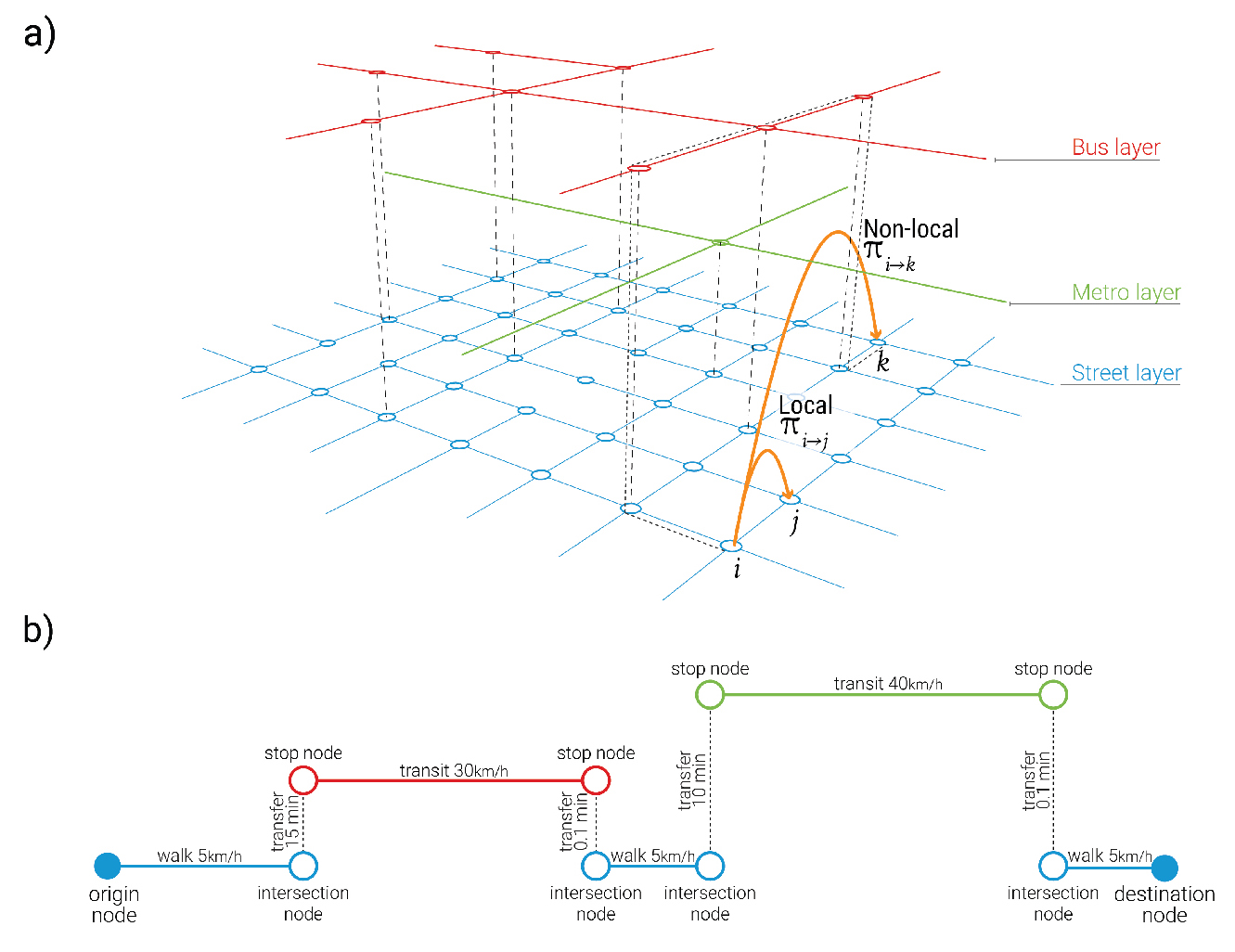}
            \caption{(a) Example of a multilayer network modelling different transport systems and their interconnections along with possible types of transitions that can be modelled using local and non-local random walks. (b) an illustration of a time-weighted path along the multilayered network.
            \label{fig::random_walk_illus}}
            \end{figure}

        This representation allows us to create time-weighted paths, as illustrated in figure \ref{fig::random_walk_illus}b,and calculate the probabilities of different nodes in the system being occupied by different population groups to measure segregation. 
        
    \subsection*{Dynamic isolation index}
    
         To calculate segregation in a city we define a transition matrix $P$ that contains entries $\pi_{i \rightarrow j}$, which indicate the probability that a random walker transitions from node $i$ to node $j$ at each time step. Additionally, we define a parameter $\alpha \in [0,1)$ that encodes the probability that at each time step the random walk continues, or stops (with probability $1-\alpha$). Given this transition matrix and the parameter $\alpha$, the probability of a walk starting in $i$ and ending in $j$ is defined as $q_{ij}$ in a matrix $Q$ such that $Q = (1-\alpha)(I - \alpha P)^{-1}P$. 
        
        Given the initial concentrations $c_{b}$ and densities $d_{b}$ of populations groups across the city,  the normalised segregation ($\bar{\sigma}_{b,i}$) of each group $b$ in the node $i$ is defined as follows:
        \begin{equation}
        \bar{\sigma}_{b,i} = (\frac{n_{b}}{n})^{-1} d_{b,i} \sum_{j} q_{ij}c_{b,j}
        \end{equation}
        where $\bar{\sigma}(b,i)$ is equal to the probability, up to the multiplicative scalar $n$, that a randomly chosen individual from group $b$ encounters another randomly chosen individual from the same group. Note that if $\bar{\sigma}(b,i) >1 $ we can say that group $b$ is isolated. The segregation index of group $b$ in the city is the average of the segregation indices of all the nodes for that group: $\bar{\sigma}_{b} = \sum_{i} \bar{\sigma}_{b,i}$, and the segregation of the city is defined as the weighted average over the segregation indices of the groups: $\bar{\sigma} = \sum_{b} \frac{n_{b}}{b} \bar{\sigma}_{b}$.
    
        Given this definition, the segregation index depends on two values: $\alpha$ and the probability of transition $\pi_{i \rightarrow j}$. $\alpha$ encodes the temporal constraints on mobility and can be related directly to the amount of time people are willing to spend on travel $\tau$, where the expected $\tau$ of a random walk given $\alpha$ is:
        \begin{equation}
        \mathbb{E}(\tau)=\overline{T_{\mathcal{M}}}\frac{1}{1-\alpha}.
        \end{equation}

\subsection*{Random walk strategies}

    Different types of random walks can be explored in terms of the weight matrices \cite{riascos2021random}. In this work we look at the diffusion process of a random walk to measure segregation by using two different strategies: local and non-local random walks. Local strategies restrict a random walker to move only to neighbouring nodes on a network. On the other hand, non-local strategies allow the random walker to move beyond the immediate neighbours. The probability of this happening is determined by a generalised cost associated with moving to a specific location as depicted in the figure \ref{fig::random_walk_illus}a. 

    \subsubsection*{Local random walks}

        \textit{Normal random walk}: In this case, the weights coincide with the elements of the adjacency matrix \cite{PhysRevLett.92.118701}, from which we can calculate the transition matrix given by :
        \begin{equation}
            \pi_{i \rightarrow j} = \frac{A_{ij}}{k_{i}}.
        \end{equation}    
        By definition, the normal random walker hops with equal probability from a node to one of its nearest neighbours in the network.
        
        \textit{Preferential navigation}: In the preferential navigation case, a random walker transitions to a neighbour with a probability that depends on a quantity $q_{i}>0$ assigned to each node $i$ of the network. The value $q_{i}$ can represent a topological feature of the respective node or a value independent of the network structure, that quantifies an existing resource at each node,
        \begin{equation}
            \pi_{i \rightarrow j} = \frac{A_{ij}q_{j}^{\beta}}{\sum_{l=1}^{N}A_{il}q_{l}^{\beta}},
        \end{equation}
    
        where $\beta$ controls the influence of the incorporated features into the random walk. Such features can encompass information about the global structure of the network, for example, in our case we select $q_{i}$ as the betweenness centrality of vertex $i$. 

    \subsubsection*{Non-local random walks}
    
        Non-local random walks on the network are motivated by the possibility of transitioning from one node to another irrespective if it is a direct neighbour. One particular type of non-local random walks are L\'evy flights. These are random walks with displacements of length $l$, and a probability distribution given by an inverse power-law relation. There have been many studies showing that human mobility patterns display this type of behaviour \cite{gallotti2016stochastic, alessandretti2018evidence, gonzalez2008understanding}.
        
        L\'evy flights on networks can be described in terms of weights $d_{ij}^{-\beta}$, where $\beta \in \mathbb{R} _{> 0}$, $\mathbb{R} _{> 0}=\left\{x\in \mathbb{R} \mid x > 0\right\}$. In the case of L\'evy flights the probability transition matrix is equal to $d_{ij}^{-\beta}$ - where the probability of transitioning from node $i$ to node $j$ in the network is a function of the time it takes to travel within the network from $i$ to $j$. We can use this probability transition matrix to calculate $Q$ such that $Q = (1-\alpha)(I - \alpha P)^{-1}P$. 

\section*{Results}

    \subsection*{Measuring segregation on an synthetic city}

\begin{figure*}[!t]
            \centering
            \includegraphics[width=0.6\linewidth]{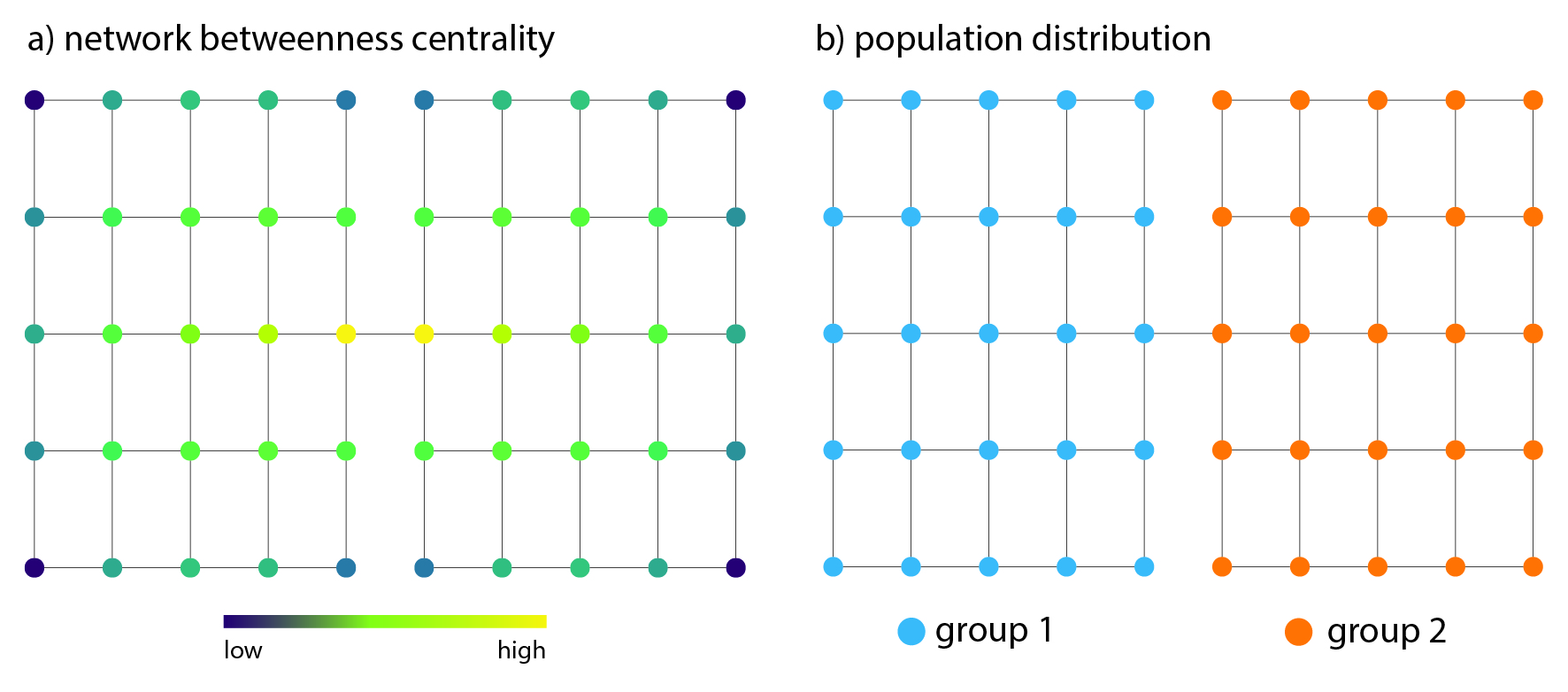}
            \caption{\label{fig:rw_ideal} Features of toy model. 
            a) Street network structure and betweenness centrality values, b) population distribution by groups, each node contains 10 people.}
            \end{figure*}

        We test the method by first running the different types of random walks and measuring the resultant segregation on a toy model represented by a mono-layered network comprised of 50 nodes and a population of 500 divided into two groups, as shown in figure \ref{fig:rw_ideal}. For this toy model, we test the three types of random walks to visualise how the network structure affects segregation, and run a sensitivity analysis for the parameters on both local and non-local random walks.
            
\begin{figure*}[!t]
            \centering
            \includegraphics[width=5in]{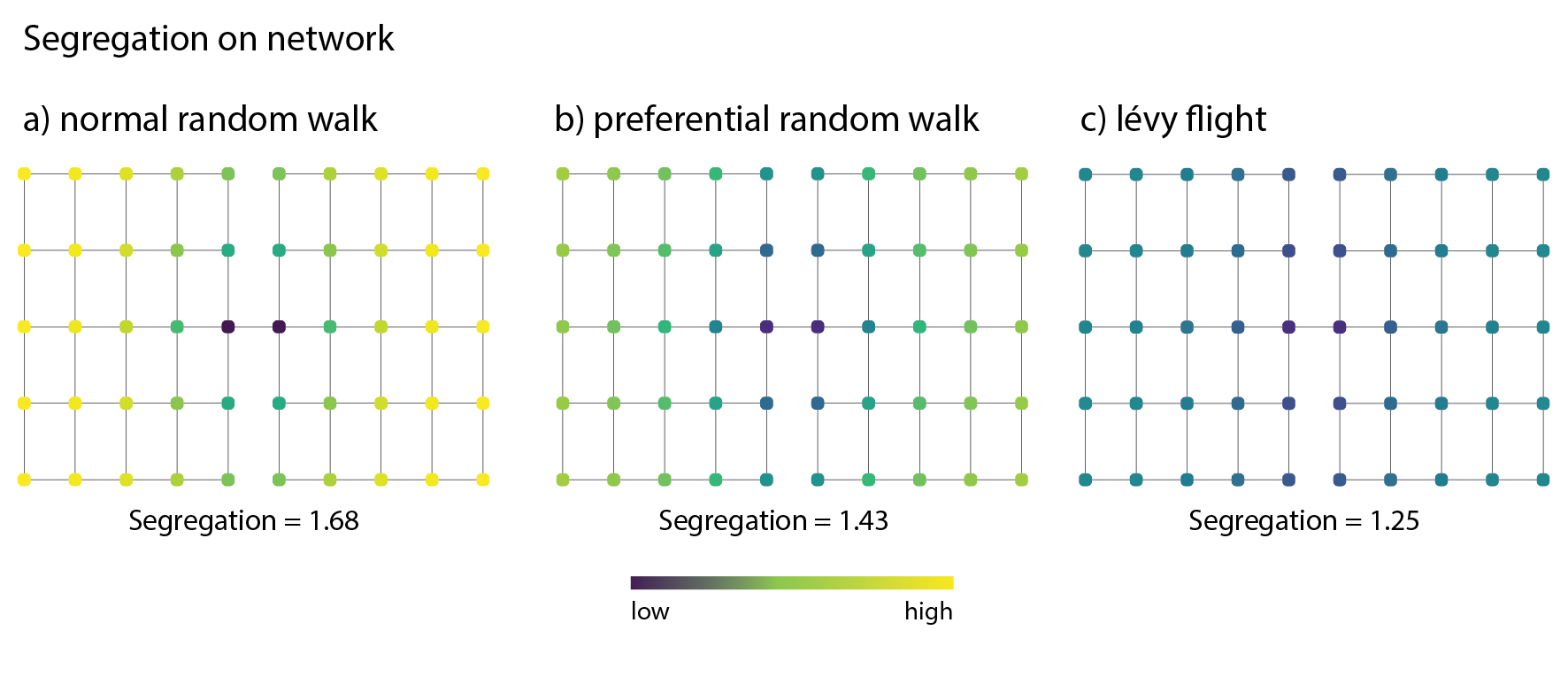}
            \caption{\label{fig:rw_result} 
            Segregation measures on the toy model city for: a) normal random walk with $\alpha=0.85$, b) preferential random walk with $\alpha=0.85$ and $\beta=1$, and c) L\'evy flight with $\alpha=0.85$ and $\beta=2$.}
        \end{figure*}

        Figure \ref{fig:rw_result} shows the influence that different dynamics have on the measure of segregation on the toy model. The measure is able to capture the influence of the network structure on segregation, where not only are the nodes that act like bridges between the two communities less segregated, but also the nodes that are within a short network distance. When the dynamics change from normal to preferential random walk, segregation decreases, as most people  will tend towards a smaller subset of final positions regardless of their initial position. In the case of a L\'evy flight, the segregation is the lowest, this is not directly comparable to the previous two types of random walks since, in this case, the $\alpha$ value no longer represents the same temporal constraint. This is because it's the relationship between $\alpha$ and expected travel time is influence by the step length distribution rather than the temporal sequence of steps. This changes the characteristics of the walk, allowing more long-distance jumps, which impacts segregation differently compared to the other two dynamics.

        \begin{figure}
            \centering
            \includegraphics[width=1\linewidth]{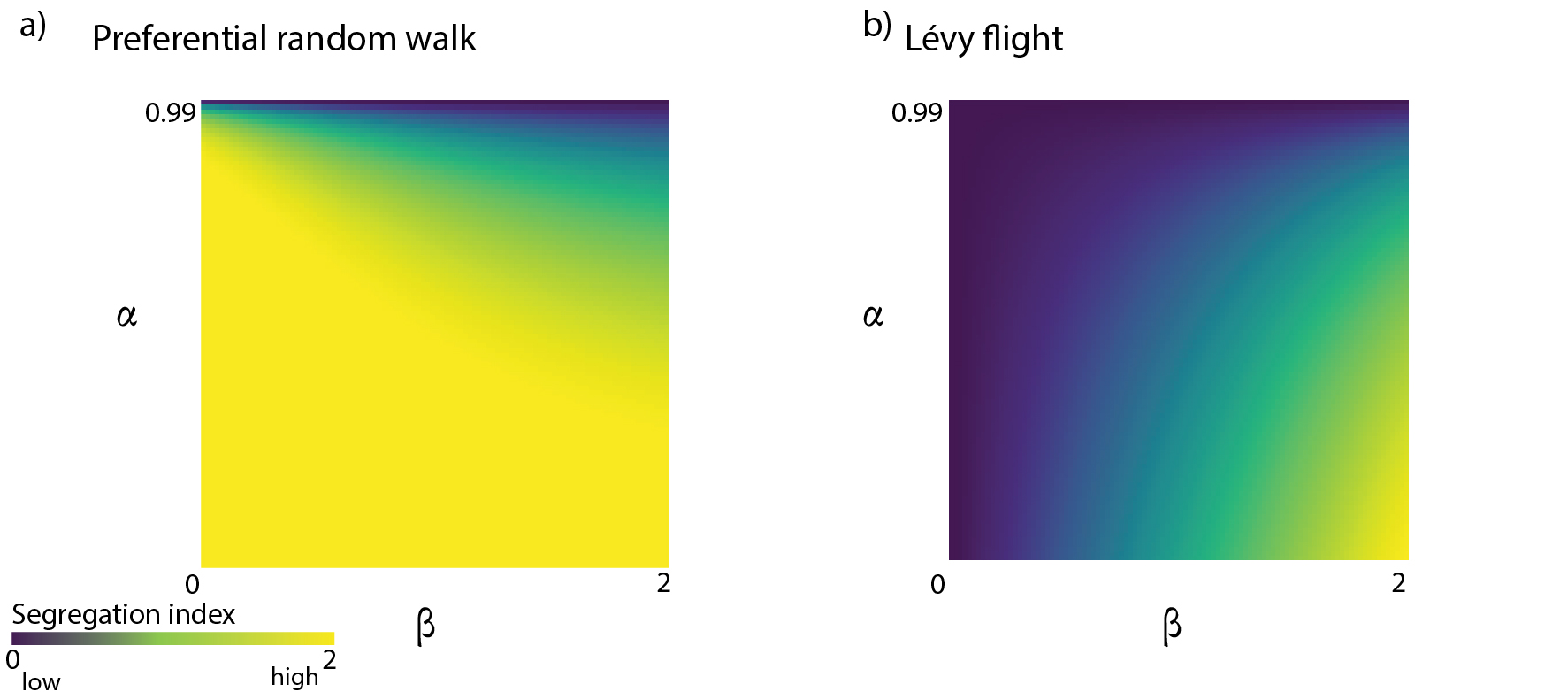}
            \caption{\label{fig:sensitivity} Sensitivity analysis for a) Local (preferential random walk) model and b) Non-local (L\'evy flight) model. In the case of local random walks the model is mainly driven by $\alpha$ - as it controls the spatial mobility of the population, with values closer to zero resulting in interactions that are constrained to direct neighbours. In the case of non-local random walks the model is mainly driven by $\beta$ - with segregation being the lowest for values close 0 where there is equal probability of transitions between nodes regardless of distance.}
        \end{figure}
    
        We perform a sensitivity analysis of both preferential and L\'evy flight models of segregation, in order to show the impact of the parameter values on the segregation index, see Figure \ref{fig:sensitivity}. It is important to note that a segregation index closer to zero indicates no segregation, while increasing values signify higher segregation levels. 
        
        In the preferential random walk model, the segregation index is primarily influenced by parameter $\alpha$; as $\alpha$ and $\beta$ increase, segregation decreases due to greater spatial mobility and individuals converging to the same subset of nodes. For the L\'evy flight model, higher values of $\beta$ lead to reduced spatial mobility, and as $\beta$ approaches two, segregation increases dramatically.
    
    \subsection*{Evaluating a transport network intervention strategy}  

        We test the proposed model on the city of Cuenca, Ecuador to investigate how the introduction of a new mode of transport affects segregation of different socio-economic groups. Cuenca, with a population of $505,585$ according to the 2010 census, is the third largest agglomeration in Ecuador after Guayaquil and Quito. It serves as a service and market centre for the southeastern region of the country \cite{bolay2004intermediate}, and the urban area extends over $72.48 Km^{2}$, housing $331,885$ people.
        
        Previous research on urban development in Cuenca which studied the evolution of city's size and population density has emphasised the importance of planned densification policies to counteract the negative impacts of urban sprawl and create liveable urban areas \cite{hermida2015densidad}. Research on socio-spatial segregation in the city, which used measures of evenness, exposure, and clustering, has shown that there are two parallel processes of spatial segregation occurring within the urban area: segregation of low socio-economic status households towards the north and west periphery of the city, and self-segregation of high socio-economic status households in areas along the Tomebamba River \cite{orellana2014segregacion}.
        
        In our study, we analyse the impact of both the bus network, and the tram which began its operation on the 25 May 2020, on Cuenca. We use the various measures described in the previous section to assess segregation in the city at a finer spatial scale by analysing population data at census block level and incorporating the spatio-temporal constraints that the various transport networks impose.
        
        To achieve this, we first classify the population into four groups based on socio-economic data from the census. We then merge this data to the street network, and build three distinct multilayered networks which contained only the street network ($\mathcal{M}{s}$), the street network and bus network ($\mathcal{M}{sb}$), and finally the street network, bus network and tram network ($\mathcal{M}{sbt}$) respectively. Finally we apply the different types of random walk segregation measures introduced previously to each network, in order to understand the effects that each has on the segregation of the different groups within the city. 
        
\begin{figure}
        \centering
        \includegraphics[width=1\linewidth]{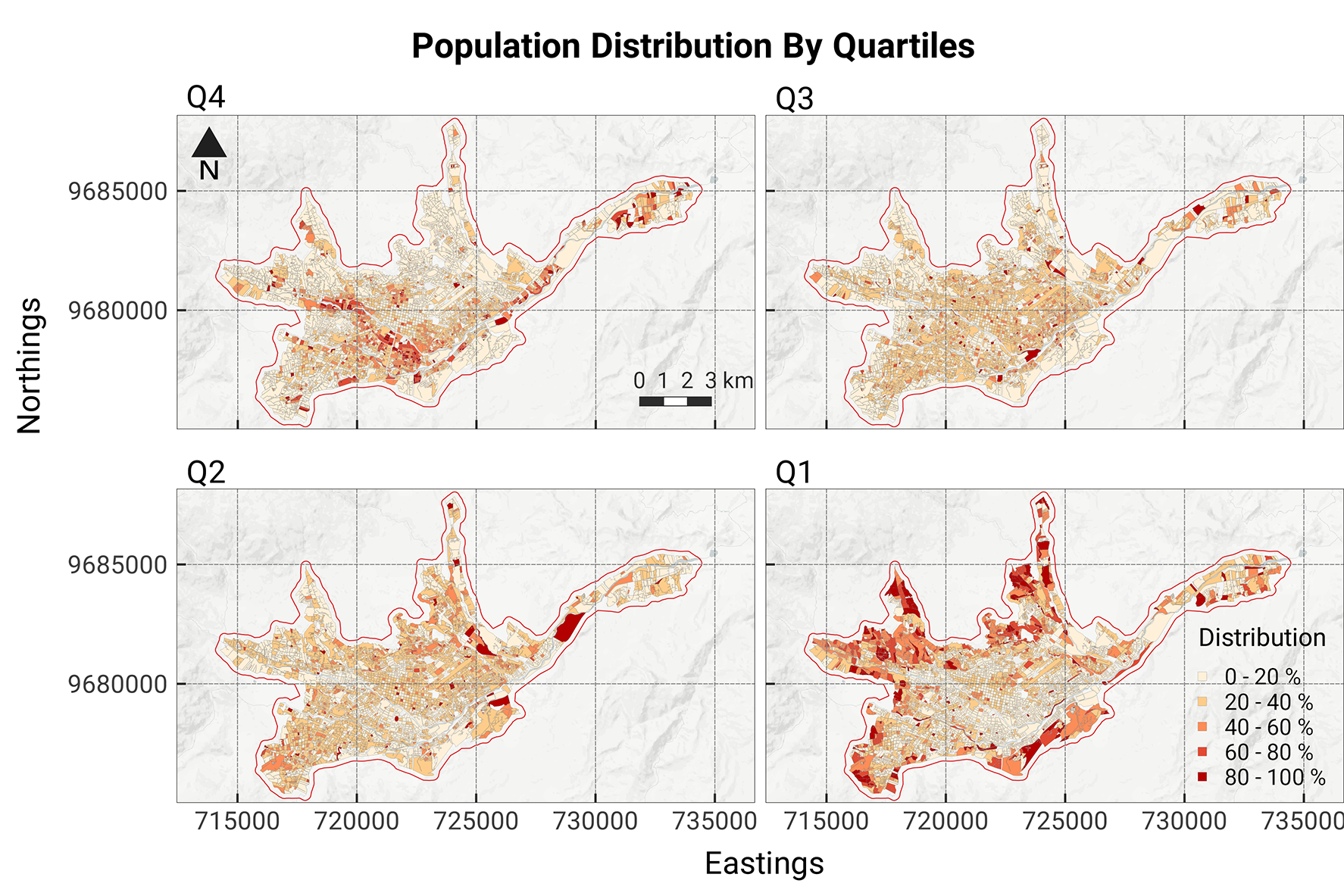}
        \caption{\label{fig:population_map}Population distribution by ICV index quartiles, with Q4 representing the highest life conditions and Q1 the lowest. Q1 clusters are primarily in the city's north and west periphery, while Q4 clusters align with the Tomebamba river on a west-east axis. Q2 and Q3 groups show a more even distribution throughout the urban area.}
        \end{figure}

        To classify the population in Cuenca into four distinct socio-economic groups we first calculate an index of life conditions \cite{orellana2014segregacion} at the household level. The index incorporates various factors such as the physical characteristics of dwellings, basic services of the household, education levels of residents, and access to health care. The index of life conditions (ICV) ranges from $0-2$ where households with less than one express deprivation and above one present life conditions above standard. The ICV values are calculated at the dwelling level, and then assigned to all individuals who reside within the dwelling. Using this index, we create four groups by dividing the resultant data into quartiles and we classify each person within the city into one of these four groups. We then aggregate the population for each group at the urban block level, resulting in a distribution of the population at a fine-spatial scale. The spatial distribution of the mean values of the index of life conditions at the block level can be found in the appendix, and the distribution of each group within the city in fig \ref{fig:population_map}.
    
        Most individuals in the city have ICV values less than one, showing that they do not meet the minimum threshold for well-being across one or more variables. The ICV values across the population follow a normal distribution with a mean of 0.9. Given this distribution the population is divided into quartiles, $B = (Q1,Q2,Q3,Q4)$. Figure \ref{fig:population_map} shows the spatial distribution of concentration $c_{b}$  of each group at the census block level. Quartiles Q1 and Q4 show clear spatial patterns, individuals with high ICV values locate along the Tomebamba River and towards the north-east while individuals with low ICV values locate towards the periphery, with some concentration within the city's historic center.
        
        Socio-economic information is obtained from census data captured at the block level. The block geometries define the urban area that will be studied. Since these geometries are not connected, their nodes are extracted as points and an alpha shape \cite{edelsbrunner1983shape} is used to determine the bounding polygon to define our study area. Different alpha values were tested to arrive at an optimal urban boundary for the case study. The resulting area is used to obtain the street network data, as well as set spatial limits on the other transport networks. 
        
        Transport network data is obtained from Open Street Map (OSM) and shapefiles provided by llactaLab - Sustainable Cities Research Group at the University of Cuenca. To construct the street network, osmnx \cite{boeing2017osmnx} is used to download and construct the network. All street segments are included, except those that relate to private streets, emergency access, steps, cycleways, and paths. Since the street network has to be modelled as a walkable layer, street directionality is disregarded by adding additional reciprocal links to all oneway streets. In addition, the distance weight attribute of the links is turned to a time-weighted attribute by multiplying the distance of each link by an average walking speed equal to $5 Km/h$.
        
        To construct the bus and tram networks similar approaches are taken. For both networks the available data consists of two shapefiles, one containing the line geometries of the transport routes and another containing point geometries of the stops or stations within the city. Custom functions are developed to transform these shapefiles into multidigraphs, which preserve information about their geometric properties.
        
        The first step involved processing the line geometries so that all routes are represented as a single polyline. Since the point geometries, representing the stops or stations do not always match the polyline geometries and contain no information about which line a specific stop belongs to additional processing is needed to match stops/stations to their routes.
        
        To address this, a $50m$ buffer is set for each route. All points which fall within this buffer are aligned or 'snapped' to the corresponding route. Once the geometries of the routes and stops/stations are matched a multidigraph for each route is created by cutting the line geometries by the points and creating the corresponding nodes and links. Geometric properties are conserved for visualisation purposes, and a temporal weight is added to each link by calculating travel time using an average travel speed of $30Km/h$ for the bus network and $40Km/h$ for the tram network. 
        
        After creating a multidigraph for each route for the bus and tram network, transfer links are created for routes within each that shared the same stop/station. These transfer links are weighted by an average waiting time of $10$ minutes for the bus network and $5$ minutes for the tram network. The resulting graphs are: 1) a strongly connected multidigraph for the street network, defined as $G_{s} = (N_{s}, L_{s}, w_{s})$; 2) a strongly connected multidigraph for the bus network, defined as $G_{b} = (N_{b}, L_{b}, w_{b})$; and 3) a strongly connected multidigraph for the tram network, defined as $G_{t} = (N_{t}, L_{t}, w_{t})$.
        Each graph can be described by their adjacency and time-weighted adjacency matrix. The topological and geometric structure of these graphs are shown in the appendix. 
        
\begin{figure*}[!t]
        \centering
        \includegraphics[width=0.8\linewidth]{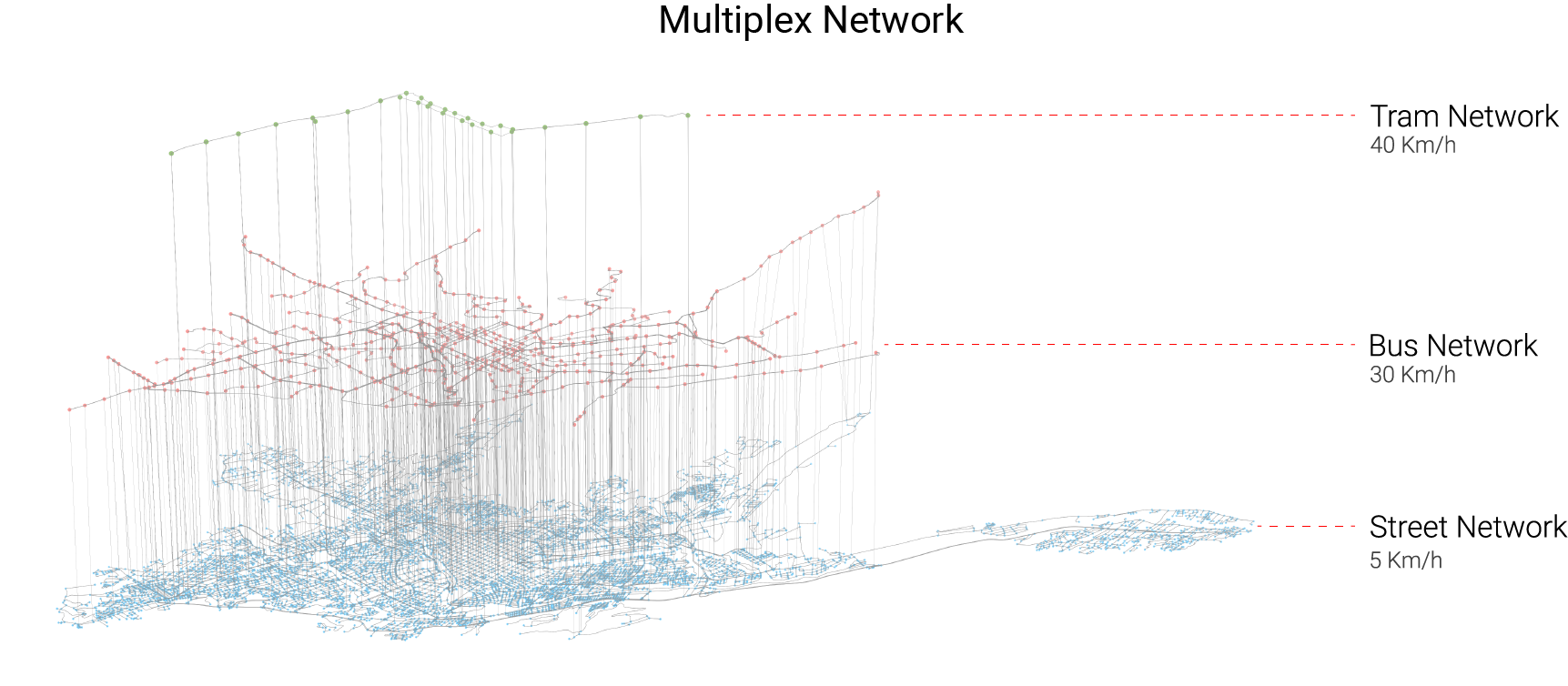}
        \caption{\label{fig:multilayered_cuenca}Multilayer network $\mathcal{M}^{sbt}$ of Cuenca showing the streets, bus, and tram networks as time-weighted graphs, along with their interlinks.}
        \end{figure*}

        Once each individual transport network is modelled as a graph, we follow the procedure described in the methodology section to create the different multilayered networks $\mathcal{M}{s}$, $\mathcal{M}{sb}$, $\mathcal{M}{sbt}$. For each multilayered network we assign the calculated socio-economic groups described previously to the nodes of the street network layer. To achieve this, we first create a Voronoi tessellation using the street intersection geometries. We then use a weighted area overlay interpolation to calculate population values for the Voronoi polygons. Finally, we assign the Voronoi polygon values to their corresponding node in the street network. This approach assumes that each individual living in a particular block will always start their journey from the same intersection, for the sake of simplicity. A representation of the multilayer network which captures all three transport modes is shown in fig \ref{fig:multilayered_cuenca}. The properties of each individual transport network, as well as each resultant multilayered network can be seen in table \ref{table:network_measures}.

        \begin{table}
        \centering
        \begin{tabular}{lrrrrrr}
        \hline 
        \vspace{-0.3cm} \\
        Network& V & E & $\bar{ \ell_{ij}}$ \tiny{Km} & $\bar{ \tau_{ij}}$ \tiny{min} & $\langle \ell \rangle$ \tiny{Km} & $\langle \tau \rangle$ \tiny{min}\\
        \hline
        $G_{s}$      & 8836    & 24554 & 6.21 & 74.55 & 23.045 & 276.54    \\
        $G_{b}$          & 1090        & 15848 & 4.79& 31.97 & 18.20 &  71.74   \\
        $G_{t}$       & 41    & 73 & 4.68& 18.28 & 11.2 & 42   \\
        $\mathcal{M}^{sb}$        & 11016     & 44762 & 5.61 & 38.31 & 22.23 & 170.72 \\
        $\mathcal{M}^{sbt}$      & 11059  &44921 & 5.60 & 37.75 & 22.23 & 170.72   \\
        \hline
        \end{tabular}
        \caption{Transport network measures of Cuenca outlining the number of nodes $N$ and links $L$ for each graph, as well as the average distance weighted shortest path $\bar{ \ell_{ij}}$ , average travel-time weighted shortest path $\bar{ \tau_{ij}}$, distance weighted diameter $\langle \ell \rangle$ and travel-time weighted diameter $\langle \tau \rangle$.}
        \label{table:network_measures}
        \end{table}
    
        We analyse segregation given the population distribution through the $\mathcal{M}_{s}$, $\mathcal{M}_{sb}$ and $\mathcal{M}_{sbt}$, to capture the effects of the spatio-temporal constraints these networks impose, and how the bus network and the  introduction of the tram affect segregation in the city. For all three networks a value of $\alpha = 0.85$ is used, which is equal to an expected weighted walk length of $20$ minutes on the street network and $25$ minutes on the multilayered networks. These values were chosen to match the mean travel time of the population as captured through travel surveys. To plot out the spatial distribution of the calculated segregation values, we aggregate the values calculated at each node of the street network to a hexagonal grid using H3 geospatial indexing system at level 9, which corresponds to hexagons that have side lengths of roughly 200m and with areas of roughly $0.1km^2$, and then we take the mean. Additionally, we run the three types of random walks described previously and highlight how each type of random walk can reveal different aspects of how the transport system affects segregation.
    
        Given that the groups Q1 and Q4 present the highest levels of spatial clustering, and have been found to have the highest levels of segregation in the city in previous studies \cite{orellana2014segregacion}, we focus on these two groups to explore the results of this study in more detail, while reporting overall values of segregation for all groups at the city level. As it will be shown, the addition of the tram does not significantly reduce segregation in the city - and change is mostly driven by considering the bus network. Because of this, we only plot the spatial distribution of change in segregation caused by considering all transport networks, and report only the city wide results for all three multilayered network models.

\begin{figure*}[!t]
        \centering
        \includegraphics[width=0.9\linewidth]{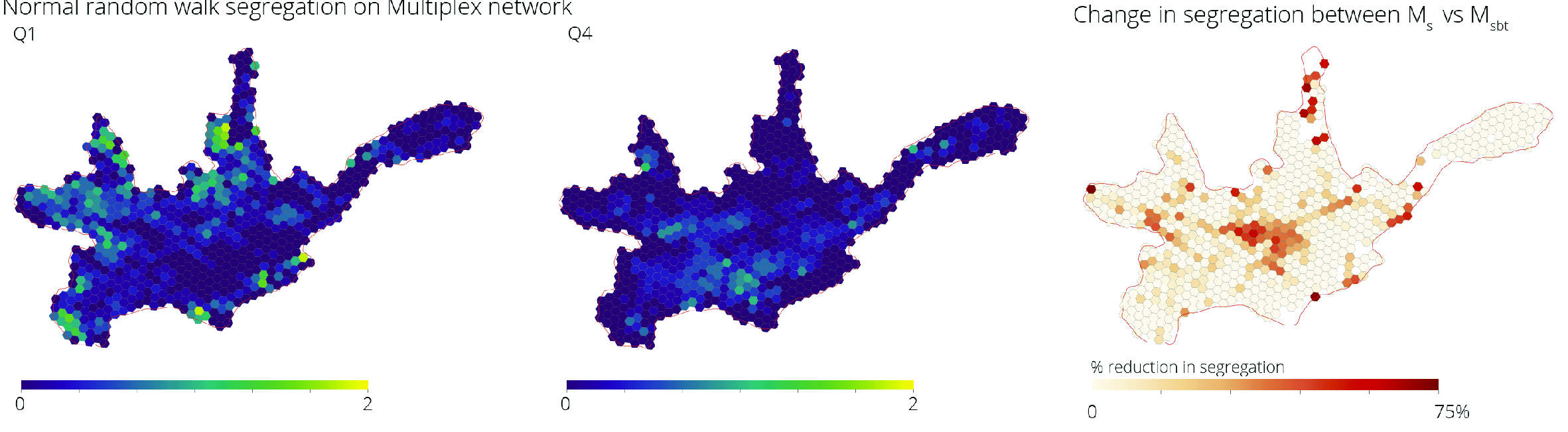}
        \caption{\label{fig:segregation_nrw}Spatial distribution of normalised segregation index for groups Q1 and Q4 in Cuenca measured using normal random walks on $\mathcal{M}^{sbt}$. Although the distribution on the Q4 population is concentrated along the east-west axis following the Tomebamba river, only those in the south-east present high levels of segregation due to a lack of transport connections to other parts of the city. Relative change in the segregation index caused by the introduction of the bus and tram network, measured using normal random walks, shows that the bus and tram reduce segregation mainly in the city centre and in specific areas in the periphery.}
        \end{figure*}

        \textbf{Normal random walk}: 

        The normal random walk segregation measure is only affected by the topological structure and connectivity of the transport networks available. 
        Fig \ref{fig:segregation_nrw} shows the resultant spatial distribution of the segregation values for groups Q1 and Q4 when considering all transport networks. Overall, the group with the higher values of segregation, Q1, corresponds to the one with the lowest index of life conditions, and this is localised mostly in the western and northern periphery of the city. Q4 also presents some spatial clustering in the southern area of the city along one of the rivers. 
        
        It is interesting to note that the segregation values that result from the normal random walk method in the street network yield similar results to the segregation values obtained through the relative size of the population groups in each census to the total size of the population group in the city. This might be due to the fact that temporal mobility of $20$ minutes introduced through $\alpha$ is similar to the group interactions that one would expect at the census track level.
        
        In fig \ref{fig:segregation_nrw} we can see the relative change in the values of the segregation index when taking into account all transport networks available, as opposed to only considering the street network. The central area of the city presents the highest decrease in segregation due to the increased connectivity to most other areas of the city through both the bus and tram network. Additionally, other areas such as specific regions in the north and west of the city also see significant reduction, and correspond to the areas in the periphery that have good coverage by the bus network. 
        
        \textbf{Preferential random walk}: We can also take into account the fact that people will tend to visit certain areas with more or less frequency depending on what those areas have to offer. In this case, we can use a preferential random walk that assigns probabilities of transition not only based on the connectivity and topological structure of the network, but also on the additional information about how attractive different places are, which can be quantified as an additional parameter $q$. 
        
\begin{figure*}[!t]
        \centering
        \includegraphics[width=0.9\linewidth]{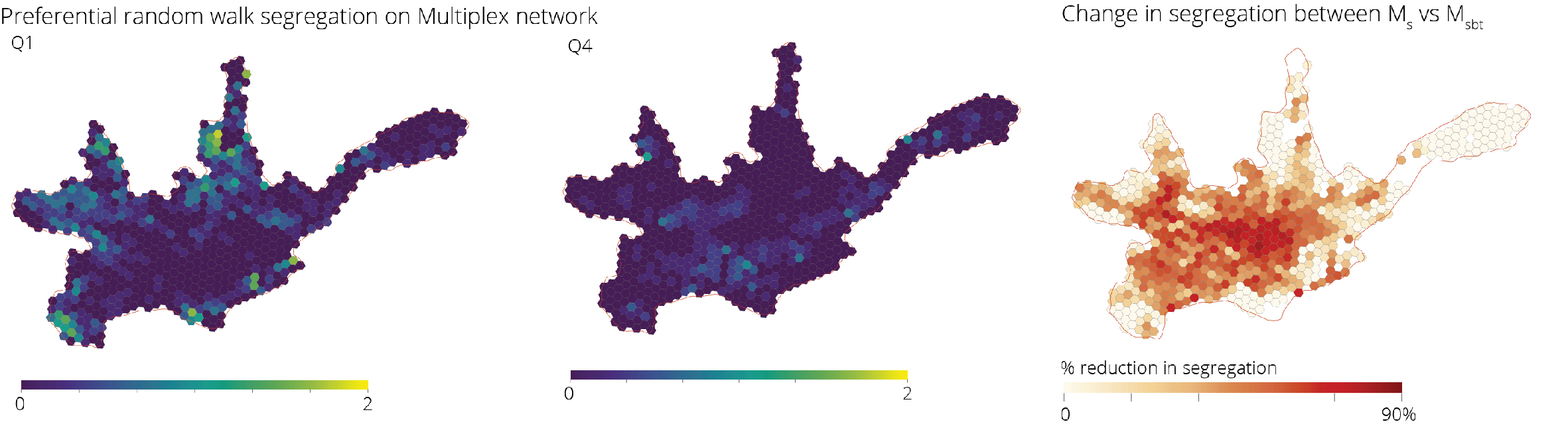}
        \caption{\label{fig:segregation_prw}Spatial distribution of the normalised segregation index for groups Q1 and Q4 in Cuenca measured using preferential random walks on $\mathcal{M}^{sbt}$. Although there is an overall lower segregation due to transition probabilities being concentrated in few places, there are still high segregation values for Q1 in the north and south-west of the city due to poor connectivity of the transport network in these areas. Relative change in the segregation index caused by the introduction of the bus and tram network, measured using preferential random walks, shows that a higher overall reduction in segregation is evident throughout most of the city, with the exception of the eastern part of the city where there is a lack of street network connectivity and no additional transport connections.}
        \end{figure*}

        For this case study, we define $q$ as the betweenness centrality of the nodes. The resultant segregation values for Q1 and Q4 can be seen in fig \ref{fig:segregation_prw}, they present similar spatial patterns of segregation to the normal random walk, with mostly a decrease of magnitude for all the areas. In this case however the reduction in segregation when considering all transport networks as opposed to only the street network, is much more pronounced, and it affects a much wider area in the city, as seen in fig \ref{fig:segregation_prw}. This is mainly driven by the fact that people will tend to visit the same areas regardless of their residence \cite{alessandretti2020scales}, increasing the probabilities that different groups end up in the same places. The areas which exhibit the highest change are areas that are well served by public transport, such as the city centre and along the linear corridors towards both the west and north of the city. 
        
\begin{figure*}[!t]
        \centering
        \includegraphics[width=0.9\linewidth]{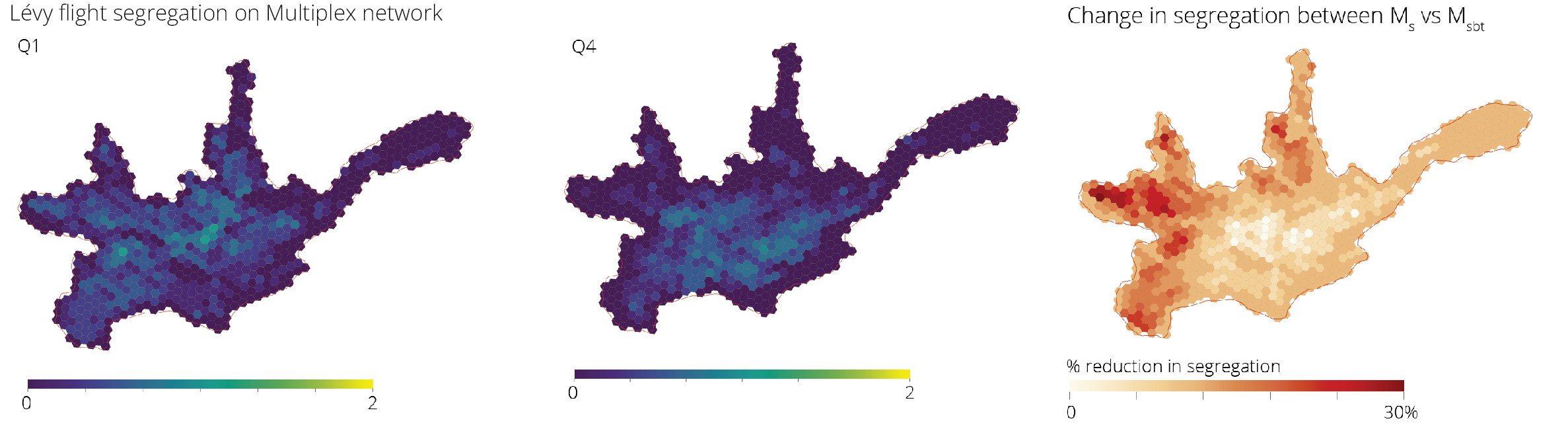}
        \caption{\label{fig:segregation_lrw}Spatial distribution of the normalised segregation index for groups Q1 and Q4 in Cuenca measured using L\'evy flights on $\mathcal{M}^{sbt}$. Segregation values tend to be less extreme overall, with higher values generally coinciding with the spatial clustering of different groups. Relative change in the segregation index, caused by the introduction of the bus and tram network and measured using L\'evy flights, shows that the tram and bus networks mostly affect the western part of the city by greatly reducing travel times towards the centre.}
        \end{figure*}

        \textbf{L\'evy flights} Finally we use L\'evy flights to better model how people move in urban areas based on how far or close different places are. In this case, the distance of the shortest paths between all nodes in the system are calculated and used to estimate a probability transition matrix $P$, this is then employed to calculate the probabilities of different individuals being present in the same area. Fig \ref{fig:segregation_lrw} shows the segregation values for Q1 and Q4 which exhibit very different spatial-patterns to both the normal and preferential random walks. Firstly, segregation values tend to be less extreme in all cases, with higher segregation values clustering near the centre and certain places in the west of the city for Q1, and towards the south east for Q4. The areas with the highest relative reduction in segregation when considering all transport modes as compared to the street network, are mostly concentrated in the west of the city as shown in fig \ref{fig:segregation_lrw}. 
        
        \begin{table*}
        \centering
        \setlength{\tabcolsep}{6pt}
        \renewcommand{\arraystretch}{1.25}
        \begin{tabular}{lrrrr|rrrr|rrrr}
        \hline
        & \multicolumn{4}{c|}{Normal} & \multicolumn{4}{c|}{Preferential} & \multicolumn{4}{c}{L\'evy Flights} \\
        Network & Q1 & Q2 & Q3 & Q4 & Q1 & Q2 & Q3 & Q4 & Q1 & Q2 & Q3 & Q4 \\
        \hline
        $\mathcal{M}_{s}$ & 1.37 & 1.07 & 1.09 & 1.42 & 1.29 & 1.04 & 1.03 & 1.31 & 1.17 & 0.91 & 0.89 & 0.99 \\
        $\mathcal{M}_{sb}$ & 1.11 & 0.85 & 0.87 & 1.18 & 0.88 & 0.65 & 0.64 & 0.84 & 1.17 & 0.89 & 0.86 & 0.94 \\
        $\mathcal{M}_{sbt}$ & 1.10 & 0.84 & 0.87 & 1.17 & 0.87 & 0.64 & 0.63 & 0.83 & 1.15 & 0.86 & 0.81 & 0.88 \\
        \hline
        \end{tabular}
        \caption{Segregation measures for Cuenca, Ecuador by ICV quartile for different network models and random walk types.}
        \label{table:segregation_measures_combined}
        \end{table*}
        
        Table \ref{table:segregation_measures_combined} shows the normalised segregation index for each quartile group for the three networks as measured using normal random walks, preferential random walks, and L\'evy flights. As mentioned in the previous section, values greater than 1 indicate that a group is over represented in a particular area. This means that the group is less exposed to other groups when considering the probabilities of two randomly chosen individuals encountering each other as they move through the city.  The value is then zero  when the probabilities match the make up of the population of the city.
        
        If we only consider the road network using normal random walks we arrive at similar results as those of Orellana and Osorio \cite{orellana2014segregacion}, where groups Q1 and Q4 present normalised segregation values $\bar{\sigma} > 1$, showing that the process of socio-spatial segregation is more pronounced in the case of individuals with the lowest and highest ICV values. In all cases we see an overall decrease in segregation when adding the bus network, and a smaller decrease when adding the tram network, showing how increased connectivity can help integrate otherwise segregated parts of the city.  
        
        The normalised segregation in the multilayered network, $\mathcal{M}_{sb}$, shows no group over-representation. Considering the city's public transportation, $\bar{\sigma}$ across all quartiles displays no group isolation, meaning buses effectively improve mobility and access across diverse city areas despite physical constraints. However, trams don't enhance this effect possibly due to similar routes with buses, serving already integrated groups.
         
        The results from the L\'evy flight method indicate that segregation is overall lower for Q4 compared to both the normal walk and preferential walk case - with only Q1 showing higher segregation values.
        We hypothesise that this is because the areas where Q4 is located in the city are much more integrated and present higher connectivity through the transport network than those areas where group Q1 tends to cluster.
        
        Through this analysis we showed that the introduction of the tram network did not have an important observed effect on increasing interactions between different groups in the city. However, the introduction of the tram presents an opportunity to restructure bus lines, reduce redundancies between these two systems, and possibly increase efficiency as well as decrease socio-economic segregation. The framework developed here, could hence provide useful information for such a restructuring, by providing a means of comparing the impact of different proposed interventions in the city.

\section*{Discussion}

In this work, we presented a framework to measure segregation beyond the residential level, by estimating the interaction probabilities for different socio-economic groups considering the available transport networks. The different transport modes and mobility constraints were modelled as random walks in a multilayered network. This method assesses the impact of new infrastructure and quantifies each transport network's contribution to the interaction opportunities. Our measure includes a parameter $\alpha \in [0,1]$, that represents the temporal constraints in urban movement. For example, applying this method to a toy model, and conducting a sensitivity analysis, we find that $\alpha=0$ results in isolation index, whereas increasing $\alpha$ captures the steady-state random walk process and the city's socio-economic distribution.

In our empirical analysis, we studied the city of Cuenca, Ecuador, using block-level socio-economic census data. We showed that the measure is able to capture the influence of the network structure on segregation, where not only are the areas of the city that act like bridges between the two communities less segregated, but also the areas that are within a short network distance, revealing patterns of segregation otherwise hidden to traditional residential segregation measures. Finally, we showed how the measure can be used to assess the introduction of new transport infrastructure, by applying the method to evaluate the introduction of a new tram to the city. Our findings show that unless the current bus network is reorganised, the effects of the new tram on integrating different socio-economic groups by increasing probabilities of interactions across different parts of the city are negligible.

There are several limitations to our study that should be considered. First, our analysis is based on a single city, so it is not clear how well the results would generalise to other cities. Second, our measure of segregation is based on random and L\'evy walks, which may not capture all the dynamics of segregation in real cities. Finally, our analysis does not take into account other factors that may influence segregation, such as cultural preferences and the distribution of amenities and jobs that act as attractors to different places.

Despite these limitations, the segregation index proposed is a valuable tool for studying the effects of the transportation network on segregation. Overall, our study provides a new perspective on the dynamics of segregation in urban environments and offers a promising approach for measuring and analysing this complex phenomenon.

\section*{Author Contributions}

\noindent
M.N. developed the theoretical framework, designed the experiments and conducted the analysis. C.M., S.M., and E.A. contributed to the theoretical framework. M.N. wrote the paper, and all authors contributed to the structure and fine tuning of the manuscript.

\begin{acknowledgments}
  This work was supported by The Alan Turing Institute’s Turing studentship scheme.
\end{acknowledgments}

\bibliography{references}

\end{document}